\documentclass{ws-procs975x65}
\def\beq{\begin{equation}}
\def\eeq{\end{equation}}
\begin{document}

\title{Microlensing events in the Galactic bulge}
\author{Mar\'ia Gabriela Navarro$^{*,1,2,3}$ Dante Minniti$^{1,3,5}$ Roberto Capuzzo-Dolcetta$^{2}$ Rodrigo Contreras Ramos$^{3,4}$  and Joyce Pullen$^{3}$
}

\address{ $^1$Departamento de Ciencias F\'isicas, Facultad de Ciencias Exactas, Universidad Andres Bello, \\
Av. Fernandez Concha 700, Las Condes, Santiago, Chile\\
$^*$E-mail: gabriela.navarro@roma1.infn.it}
\address{$^2$Dipartimento di Fisica, Universit\`a degli Studi di Roma ``La Sapienza'', \\ P.le Aldo Moro, 2, I00185 Rome, Italy}
\address{$^3$Millennium Institute of Astrophysics, \\ Av. Vicuna Mackenna 4860, 782--0436, Santiago, Chile}
\address{$^4$Instituto de Astrofisica, Pontificia Universidad Cat\'olica de Chile, \\ Av. Vicuna Mackenna 4860, 782--0436 Macul, Santiago, Chile }
\address{$^5$Vatican Observatory, \\ V00120 Vatican City State, Italy }

%\author{Anthony N. Author} 
%\address{Group, Laboratory, Street,\\
%City, State ZIP/Zone, Country\\
%E-mail: an\_author@laboratory.com}

\begin{abstract}
For the first time we detected microlensing events at zero latitude in the Galactic bulge using the VISTA Variables in the V\'ia L\'actea Survey (VVV) data. \cite{dante} We have discovered a total sample of $N = 630$  events within an area covering $20.7$ sq.~deg.\cite{g17,g18}  
Using the near-IR color magnitude diagram we selected $N = 291$ red clump sources, allowing us to analyse the longitude dependence of microlensing across the central region of the Galactic plane. We thoroughly accounted for the photometric and sampling efficiency.
The spatial distribution is homogeneous, with the number of events smoothly increasing toward the Galactic center. We find a slight asymmetry, with a larger number of events toward negative longitudes than positive longitudes, that is possibly related with the inclination of the bar along the line of sight.
We also examined the timescale distribution which shows a mean on $17.4\pm1.0$ days for the whole sample, and 
$20.7\pm1.0$ for the Red Clump subsample
\end{abstract}

\bodymatter
\section{Introduction}
%\subsection{Subsections only have the first letter of the entire title capitalized}
%\section{Brief Comments}
Gravitational microlensing is a geometrical effect related to the apparent increase in the brightness of a background source by an object (lens) located sufficiently close to the line of sight. \cite{pac} This effect is useful to detect objects independently of their intrinsic brightness, with a wide range of masses from planets to black holes.

The main surveys dedicated to detect microlensing events in the galactic bulge are the Massive Astrophysical Compact Halo Objects (MACHO), \cite{al93} the Optical Gravitational Lensing Experiment (OGLE), \cite{u93}  the Microlensing Observations in Astrophysics (MOA), \cite{b01}  the Exp\'erience pour la Recherche d'Objets Sombres (EROS), \cite{a93} the Disk Unseen Objects (DUO), \cite{a95} the Wise Observatory \cite{y12} and the Korean Microlensing Telescope Network  (KMTNet). \cite{k10,k17}

Although a large number of events have been discovered, it is not possible to study the plane of the Milky Way ($|b|<2^\circ$) with these optical surveys due to the high extinction and crowding. However, the Galactic plane is especially interesting because, as the density of stars increases, we expect to find microlensing events in large quantities. \cite{luk}

The only way to study the plane of the Galaxy is through observations in infrared bands. Motivated by this, we carried out the first systematic search of microlensing events at low latitudes using the \emph{VISTA Variables in the V\'ia L\'actea Survey} (VVV). \cite{dante} We analysed 14 VVV tiles (from b327 to b340) covering a large area within  $-10.00^\circ \leq l \leq 10.44^\circ$ and $ -0.46^\circ \leq b \leq 0.65^\circ$. The 3 innermost tiles (b332, b333 and b334) were published in Ref.~\refcite{g17}. The other 11 tiles are presented in Ref.~\refcite{g18}.

\begin{figure}[t]
\begin{center}
\includegraphics[width=5in]{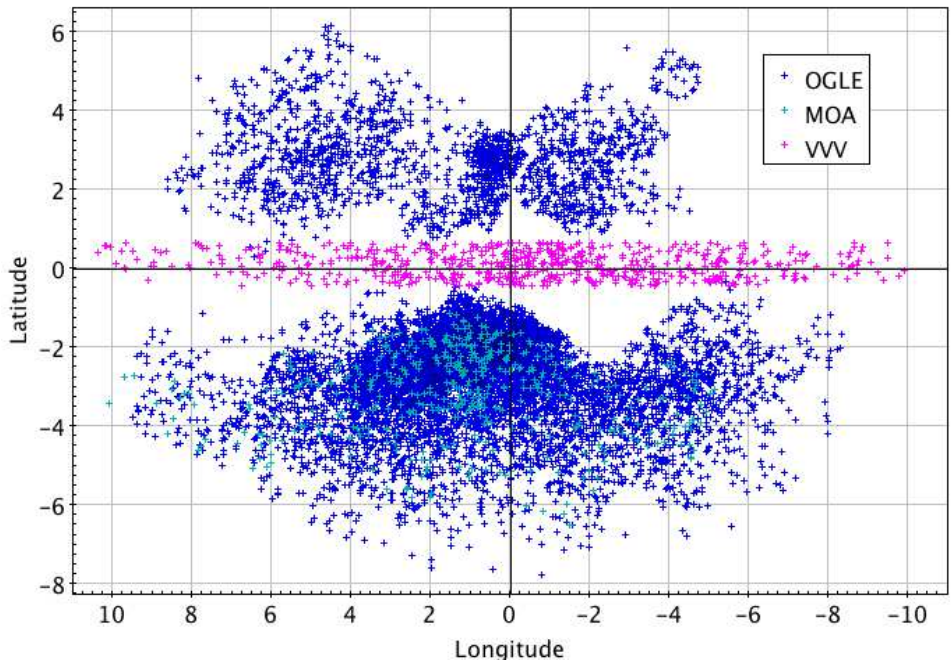}
\end{center}
\caption{Spatial distribution of the VVV microlensing events presented in this work (magenta crosses) around the Galactic center. The blue crosses are the microlensing events from the OGLE Early Warning System (EWS, http://ogle.astrouw.edu.pl) and the cyan crosses are the microlensing events discovered by MOA. All of the events happened between 2010 and 2015. Figure from Ref.~\protect\refcite{g18} reproduced with permission.}
\label{spatial}
\end{figure}

\section{The VVV data}
The \emph{VISTA Variables in the V\'ia L\'actea Survey} (VVV) \cite{dante} is an ESO public survey scanning the Milky Way bulge and adjacent section of the southern mid-plane in the near-infrared. Using the Visible and Infrared Survey Telescope for Astronomy (VISTA) 4-meter telescope located at ESO Cerro Paranal Observatory in Chile.
The area studied comprises $63 \times 10^6$ light curves for individual point sources and a multi-epoch campaign in the $K_s$-band  ($K_s<17.5$mag) with $\sim 90$ epochs spanning six seasons of observations (2010--2015). 
The PSF photometry was carried out using DAOPHOT II/ALLSTAR package. \cite{rod} The search procedure was done using the standard microlensing model, i.e., considering sources and lenses as point-like objects. 

\section{VVV Microlensing events}
\begin{figure}[t]
\begin{center}
\includegraphics[width=5.3in]{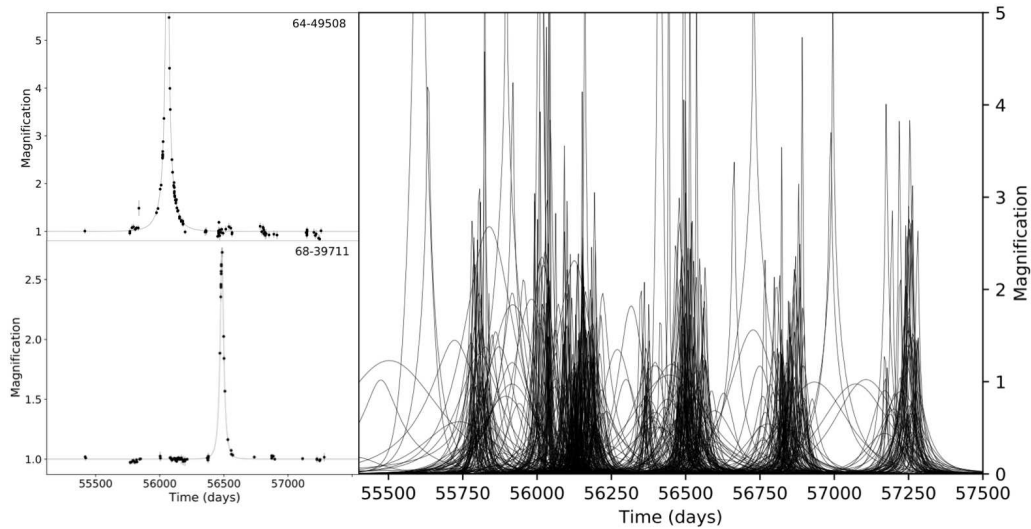}
\end{center}
\caption{
Left: Examples of two typical microlensing event light curves. Right: Fits to all 630 microlensing events discovered between 2010 and 2015. 
 }
\label{all}
\end{figure}

Initially we found 182 microlensing events in the three innermost tiles (b332, b333 and b334), with an excess in the number of events in the central tile and a relatively large number of long timescale events (with $t_E>100$ days). We extended the sample in order to study in greater detail the behaviour along the Galactic longitude. The final sample consists of 630 microlensing events within $-10.00^\circ \leq l \leq 10.44^\circ$ and $ -0.46^\circ \leq b \leq 0.65^\circ$. 
The complete sample is shown in Fig.~\ref{spatial} along with the OGLE and MOA events discovered during the same period. Fig.~\ref{all} shows all of the microlensing fits as well as two typical light curves. The animation that shows the microlensing events detected during the operation of the VVV survey is available online.\footnote{\url{https://www.youtube.com/watch?v=Uv08EAKVSQo}}
 In the upper panel an image of the studied area is shown and the flash corresponding to each microlensing event detected in their respective positions. The intensity of the flash is associated with the amplitude. In the lower panel this figure is reproduced as a function of time. The different colors correspond to the tile of each event and therefore are related to its position.

The events show a homogeneous distribution, smoothly increasing in numbers towards the Galactic centre, as predicted by different models due to the high stellar density. Considering that we expect to find a smaller amount of microlensing events due to the extreme extinction and crowding, \cite{elena} the increase in the number of events towards the center of the galaxy is expected to be even more peaked. Additionally we found an asymmetry in the density distribution with a higher number of events ($\sim60\%$) at negative Galactic longitudes, this can be explained by the inclination of the bar along the line of sight. 

\begin{figure}[b]
\begin{center}
\includegraphics[width=5.3in]{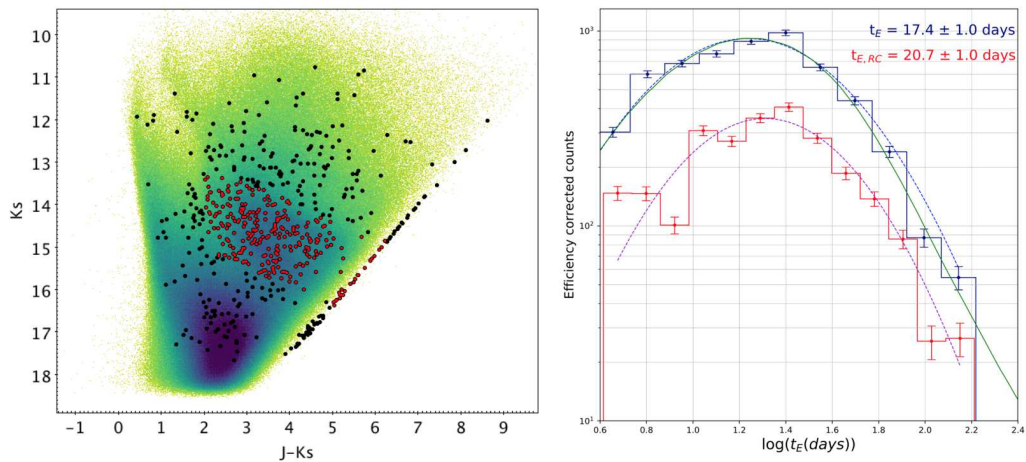}
\end{center}
\caption{
Left: $K_s$ vs.\ $J$-$K_s$ color magnitude diagram for the 14 VVV tiles. 
The black circles are the microlensing sources, and the red circles are the red clump sources.
Right: Efficiency corrected timescale distributions. The blue histogram is the distribution of the complete sample. The red histogram corresponds to the RC sources. Dashed lines show the best fit model of each histogram. The green shows the model proposed by Wegg et al. \cite{wegg} normalised to the peak of our distribution. Figure from Ref.~\protect\refcite{g18}.}
\label{te}
\end{figure}

Red Clump (RC) sources are low-mass stars that are burning Helium in the core. The study of the sources located in the RC is more reliable because they are more localised in the Color-Magnitude Diagram (CMD) and we can assume that they belong to the bulge. Consequently, using the near-IR Color-Magnitude Diagram we selected the Red Clump sources to analyze the longitude dependence of microlensing across the central region of the Galactic plane. Fig.~\ref{te} shows the CMD of the 14 VVV tiles along with the final sample of events. The red circles are the events located in the RC, corresponding to $46 \%$ of the sample ($N=291$ events). \\

In order to study the timescale distribution it is crucial to implement an efficiency analysis. The photometric efficiency was performed using artificial star simulations for the RC stars. \cite{elena} The incompleteness is severe as we approach to the Galactic center, reaching values close to $40\%$. The sampling efficiency is cadence and timescale dependent, and was computed using Monte Carlo simulations for each fixed representative timescales with random impact parameters and times of maximum magnification. Due to the different cadence among the 14 tiles studied we performed the efficiency for each tile separately. With this analysis we conclude that the VVV Survey is limited by the sampling efficiency and is more efficient discovering long timescale events than short timescale events due to the large time coverage but low and irregular cadence. 

The efficiency corrected timescale distribution is shown in Fig.~\ref{te} for the complete sample and for the RC sources. The mean timescale is $17.4\pm1.0$ days for the complete sample and $20.7\pm1.0$ days for the RC sources, in agreement with previous results. We also examined the timescale distribution for different longitudes, obtaining a clear decrease in the mean timescales as we approach to zero longitudes. A similar trend was previously reported by Wyrzykowski et al. 2015\cite{luk} but at higher latitudes using OGLE data.

\section{Summary}
We detected $N=630$ microlensing events in the plane of the Galaxy using the VVV data. We performed for the first time a longitude analysis across the central Galactic plane at $b=0 \deg$. The spatial distribution shows a peak in the center as expected, although it is asymmetric, with more events towards negative longitudes. 

The efficiency analysis demonstrates that the VVV survey is successful in detecting long/intermediate timescale events in highly reddened areas. Short timescale events are harder to find due to the low cadence of the survey (nightly at best). Our results can be complemented with other near-IR surveys to study the event rate, such as the UKIRT microlensing survey, \cite{y17}  which is currently mapping the inner bulge with a higher cadence.

We computed the timescale distribution corrected by efficiency for the complete sample and RC sources. The distribution shows a peak consistent with previous measurements at higher latitudes in the bulge. 

This study has many applications, from the study of the general behaviour of the complete microlensing population, to the study of specific events such as the events showing parallax effect, binary sources or lenses, as well as the very long timescale events which favor massive lenses (like black holes \cite{dantebh}).  Furthermore, the analysis at very low latitudes is useful to optimise the observational campaign for the Wide Field Infrared Survey Telescope (WFIRST) \cite{green12} and to complement other recent near IR surveys. \cite{y17}

In this context, the study of a larger area is warranted to increase the statistics and in turn cover common areas with OGLE and MOA in order to complete a panoramic coverage of the event rates in the inner regions of the Milky Way.

%\section*{Acknowledgments}
%We gratefully acknowledge data from the ESO Public Survey program ID 179.B-2002 taken with the VISTA telescope, and products from the Cambridge Astronomical Survey Unit (CASU). \\

\end{document}